\documentclass[conference]{IEEEtran}

\usepackage{amsmath,amssymb,amsfonts}
\usepackage{graphicx}
\usepackage{booktabs}
\usepackage{threeparttable}
\usepackage{tabularx}
\usepackage{multirow}
\usepackage{xcolor}
\usepackage{soul}
\usepackage{enumitem}
\usepackage{float}
\usepackage{afterpage}
\usepackage{placeins}
\usepackage{adjustbox}

\usepackage[T1]{fontenc}
\usepackage[utf8]{inputenc}
\usepackage{microtype}

\usepackage{newunicodechar}
\newunicodechar{≥}{$\geq$}

\usepackage[caption=false,font=footnotesize]{subfig}
\usepackage{subcaption}

\usepackage{tikz} 
\usepackage[framemethod=TikZ]{mdframed}
\usepackage[most]{tcolorbox}
\usepackage{multicol}

\usepackage{pifont}
\newcommand{\xmark}{\ding{55}}

\usepackage{cite}

\usepackage{algorithm}            
\usepackage[noend]{algpseudocode} 

\algrenewcommand\algorithmicrequire{\textbf{Input:}}
\algrenewcommand\algorithmicensure{\textbf{Output:}}

\usepackage{xurl}
\usepackage[hidelinks]{hyperref}

\IEEEoverridecommandlockouts

\def\BibTeX{{\rm B\kern-.05em{\sc i\kern-.025em b}\kern-.08em
    T\kern-.1667em\lower.7ex\hbox{E}\kern-.125emX}}
    
\begin{document}

\title{RPKT: Learning What You Don't Know - Recursive Prerequisite Knowledge Tracing in Conversational AI Tutors for Personalized Learning\\}

\author{%
  \IEEEauthorblockN{Jinwen Tang\textsuperscript{*}}%
  \IEEEauthorblockA{%
    \textit{EECS Department}\\%
    University of Missouri\\
    Columbia, MO, USA\\
    jt4cc@umsystem.edu}
  \and
  \IEEEauthorblockN{Qiming Guo\textsuperscript{*}}%
  \IEEEauthorblockA{%
    \textit{Department of Computer Science}\\
    Texas A\&M University–Corpus Christi\\
    Corpus Christi, TX, USA\\
    qguo2@islander.tamucc.edu}
  \and
  \IEEEauthorblockN{Zhicheng Tang}%
  \IEEEauthorblockA{%
    \textit{Independent Researcher}\\%
    University of Missouri\\
    Columbia, MO, USA\\
    robert.tang.robot@gmail.com}%
\and
\IEEEauthorblockN{Yi Shang}%
  \IEEEauthorblockA{%
    \textit{EECS Department}\\%
    University of Missouri\\
    Columbia, MO, USA\\
    shangy@umsystem.edu}%
  \thanks{\textsuperscript{*}\,The first two authors contributed equally to this work.}%
}

\maketitle

\begin{abstract}
Educational systems often assume learners can identify their knowledge gaps, yet research consistently shows that students struggle to recognize what they don't know they need to learn—the "unknown unknowns" problem. This paper presents a novel Recursive Prerequisite Knowledge Tracing (RPKT) system that addresses this challenge through dynamic prerequisite discovery using large language models. Unlike existing adaptive learning systems that rely on pre-defined knowledge graphs, our approach recursively traces prerequisite concepts in real-time until reaching a learner's actual knowledge boundary. The system employs LLMs for intelligent prerequisite extraction, implements binary assessment interfaces for cognitive load reduction, and provides personalized learning paths based on identified knowledge gaps. Demonstration across computer science domains shows the system can discover multiple nested levels of prerequisite dependencies, identify cross-domain mathematical foundations, and generate hierarchical learning sequences without requiring pre-built curricula. Our approach shows great potential for advancing personalized education technology by enabling truly adaptive learning across any academic domain.
\end{abstract}


\section{Introduction}

Effective education requires alignment between what instructors teach and what students need to learn. However, this alignment faces some fundamental challenges that create persistent barriers to learning success.

First, the expert blind spot phenomenon causes instructors who have mastered a domain to unconsciously skip over foundational concepts that novices need \cite{nathan2001expert}. Experts develop automated knowledge structures that make it difficult to remember the incremental steps required for initial understanding \cite{hinds1999curse}. This cognitive gap between expert instructors and novice learners results in curricula that assume prerequisite knowledge students may not possess.

Second, learners face dual cognitive challenges. They suffer from "unknown unknowns", the inability to recognize knowledge gaps they need to address \cite{kruger1999unskilled}. Research consistently shows that students overestimate their understanding and cannot identify missing prerequisites \cite{dunning2011dunning}. Additionally, previous research reveals that students rate passive learning experiences more favorably despite achieving lower learning outcomes, with this preference driven by the increased cognitive effort required for active engagement \cite{deslauriers2019measuring}. The cognitive load of formulating questions about concepts they don't understand creates a barrier that prevents students from seeking necessary clarification.

Current educational technologies fail to bridge this expert-novice gap effectively. Traditional approaches rely on static prerequisite structures that cannot adapt to individual knowledge profiles \cite{bloom19842}. While intelligent tutoring systems have shown promise, they require extensive domain modeling and pre-built knowledge graphs \cite{brusilovsky2001adaptive}. Recent LLM-based educational tools can generate explanations but still assume students can articulate what they need to learn, which is an assumption invalidated by the unknown unknowns problem.

This paper presents \textbf{RPKT - Recursive Prerequisite Knowledge Tracing}, a novel system that addresses these challenges through three key innovations:
\begin{enumerate}
    \item It eliminates the cognitive burden of question formulation through binary knowledge assessment (know/don't know), enabling learners to focus on self-evaluation rather than articulation.
    \item It dynamically discovers prerequisite dependencies through recursive LLM-powered tracing, systematically revealing the "unknown unknowns" that learners cannot identify independently.
    \item It generates personalized learning paths from identified knowledge boundaries to target concepts, bridging the expert-novice gap without requiring pre-built curricula or domain-specific knowledge graphs.
\end{enumerate}
 Our approach transforms the learning problem from "what questions should I ask?" to simple binary decisions, aligning with learners' cognitive preferences while systematically uncovering unknown unknowns.

\section{Related Work}

\paragraph{Knowledge Tracing and Adaptive Learning}
Traditional knowledge tracing models focus on estimating what students know rather than discovering what they need to learn. Deep Knowledge Tracing \cite{piech2015deep} uses recurrent neural networks to model student knowledge states, while subsequent work has incorporated attention mechanisms \cite{ghosh2020context}. However, these approaches require predefined skill sets and cannot discover unknown prerequisites dynamically. Adaptive learning systems like those reviewed by \cite{brusilovsky2001adaptive} personalize content delivery but remain constrained by expert-authored knowledge structures that may not reflect individual learning needs.

\begin{figure*}[h!]
\centerline{\includegraphics[width=\textwidth]{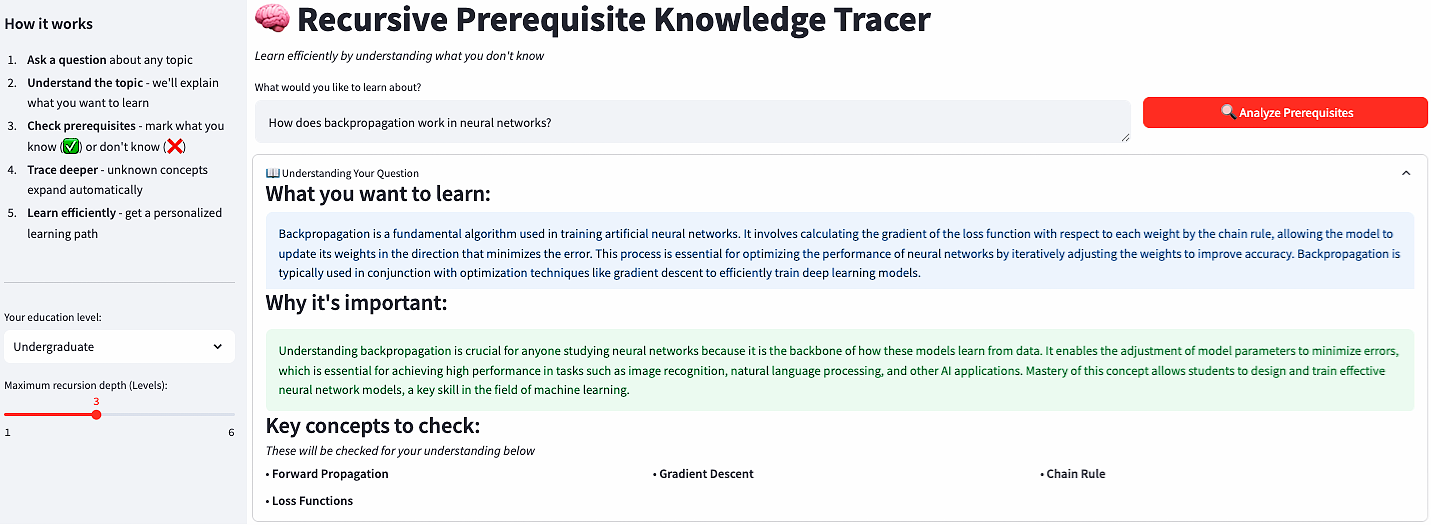}}
\caption{RPKT system interface showing initial assessment for a backpropagation query.}
\label{fig:main_interface}
\end{figure*}

\begin{figure*}[h!]
\centerline{\includegraphics[width=\textwidth]{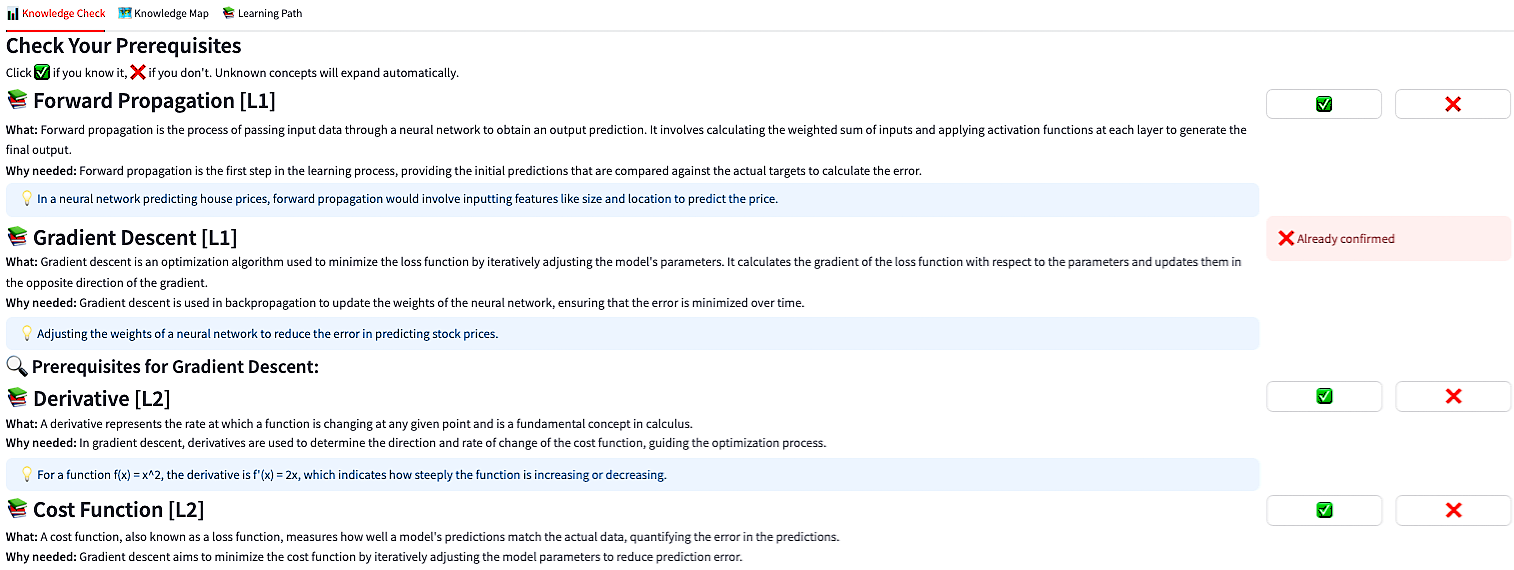}}
\caption{Example of RPKT's recursive expansion in action. After marking "Gradient Descent" as unknown at Level 1, the system expands inline to reveal Level 2 prerequisites (Derivative, Cost Function), demonstrating the recursive prerequisite discovery process.}
\label{fig:recursive_expansion}
\end{figure*}

\paragraph{Question Generation and Student Engagement}
Research on educational question generation has primarily focused on creating assessment items rather than discovering knowledge gaps \cite{kurdi2020systematic}. Studies consistently show that students struggle with question formulation, with many being reluctant to reveal knowledge gaps \cite{karabenick2003seeking}. The cognitive load theory explains why students prefer passive reception—question generation requires significant mental effort that many learners find overwhelming \cite{sweller2011cognitive}. Our binary assessment approach directly addresses this limitation by eliminating the question generation burden.

\paragraph{Recent Trends in LLM Disciplinary Applications}
Since 2023, AI has evolved from computer vision and machine learning to NLP and toward AGI through LLMs. GPT-based models have integrated into daily life while excelling in interdisciplinary domains including knowledge question-answering, medical diagnosis, mental health support, and autonomous driving~\cite{guo2024soullmate,guo2024soullmate_2,tang2025layered,tang2024advancing,yan2024llm}. LLMs have also demonstrated significant impact in clinical medicine with models like ChatDoctor providing medical consultation~\cite{li2023chatdoctor} and Med-PaLM achieving expert-level performance on medical exams~\cite{singhal2023medpalm}, legal document analysis~\cite{cui2023chatlaw}, scientific research acceleration~\cite{taylor2022galactica}, code generation and software engineering~\cite{chen2021codex}, financial market analysis~\cite{wu2023bloomberggpt}, and multimodal biomedical applications~\cite{moor2023medpalm}. These technologies increasingly automate complex cognitive tasks across diverse professional domains.

\paragraph{LLMs in Education}
Recent work has explored using large language models specifically for educational applications. ChatGPT and similar models have been used for tutoring, feedback generation, and content creation \cite{kasneci2023chatgpt}. However, these applications require students to formulate effective prompts to get useful responses, which is a significant barrier based on the difficulty in question generation. Additionally, existing LLM tools typically provide direct answers or explanations, assuming students can articulate their learning needs. \cite{abdelrahman2023knowledge} provides a comprehensive survey of knowledge tracing approaches but notes the limitation that current systems cannot identify what students don't know they don't know. Our work differs by using LLMs for dynamic prerequisite discovery rather than direct instruction, eliminating the need for students to formulate questions.



\paragraph{Prerequisite Learning and Curriculum Design}
Traditional curriculum design relies on expert-defined prerequisite chains that may not match individual learning trajectories \cite{tucker2003model}. Research in computer science education has shown that students often lack unexpected prerequisites, particularly mathematical foundations \cite{robins2003learning}. Recent work on prerequisite relation extraction from educational materials \cite{liang2018investigating} still requires existing content and cannot generate prerequisites dynamically. Our recursive approach discovers these hidden dependencies in real-time without predefined structures.

The key limitation across all related work is the assumption that either experts can fully specify prerequisite structures or students can identify their own knowledge gaps. RPKT addresses both limitations through dynamic discovery and binary assessment, creating a novel approach to personalized learning.

\begin{algorithm}
\caption{Recursive Prerequisite Knowledge Tracing}
\label{alg:rpkt}
\begin{algorithmic}[1]
\Require User question $Q$, education level $E$, max depth $d_{\max}$
\Ensure Knowledge tree $T$, explanation $\mathcal{E}$
\State $\text{analysis} \gets \textsc{AnalyzeQuestion}(Q, E)$
\State $\textit{status} \gets \{\}$, $T \gets \textsc{InitTree}(Q)$
\For{$c \in \text{analysis.key\_concepts}$}
    \State $\textsc{RecursiveTrace}(c, 1, d_{\max}, T, \textit{status})$
\EndFor
\State $\mathcal{E} \gets \textsc{GenerateExplanation}(Q, \textit{status})$
\State \textbf{return} $T, \mathcal{E}$
\Function{RecursiveTrace}{$c, \text{depth}, d_{\max}, T, \textit{status}$}
    \If{$\text{depth} > d_{\max} \lor \textsc{IsFundamental}(c)$}
        \State \textbf{return}
    \EndIf
    \If{$c \notin \textit{status}$}
        \State $\textit{status}[c] \gets \textsc{UserAssess}(c)$
    \EndIf
    \If{$\textit{status}[c] = \text{False}$}
        \State $\textsc{AddToTree}(T, c, \text{depth})$  
        \State $\text{prereqs} \gets \textsc{ExtractPrereqs}(c)$
        \For{$p \in \text{prereqs}$}
            \State $\textsc{RecursiveTrace}(p, \text{depth}+1,$
            \Statex \hspace{3em} $d_{\max}, T, \textit{status})$
        \EndFor
    \EndIf
\EndFunction
\end{algorithmic}
\end{algorithm}

\section{METHOD}

\paragraph{System Architecture}

Our Recursive Prerequisite Knowledge Tracing (RPKT) system employs a three-component architecture that discovers knowledge dependencies dynamically without requiring pre-built curricula. The Knowledge Tracer Engine leverages GPT-4o to extract prerequisite relationships in real-time based on the learner's query and educational context. This engine interfaces with an Interactive Assessment Interface presenting binary evaluations through a three-tab design. The Session Management System maintains state consistency across recursive expansion levels, tracking assessed concepts and managing prerequisite exploration depth.

Unlike traditional adaptive learning systems dependent on expert-authored knowledge graphs, RPKT discovers dependencies dynamically, enabling application across any academic domain. The system transforms complex self-assessment into simple binary decisions, addressing the cognitive burden that prevents learners from identifying their own knowledge gaps.

\paragraph{Recursive Prerequisite Extraction Algorithm}
Algorithm~\ref{alg:rpkt} presents our recursive extraction process that systematically uncovers knowledge dependencies until reaching the learner's actual knowledge boundary. Given a target concept $C_0$, the system extracts immediate prerequisites $P_1 = \{p_1, p_2, \ldots, p_n\}$ through structured GPT-4o prompting that emphasizes directly relevant technical dependencies. Each prerequisite undergoes binary assessment where learners indicate their knowledge state as $K(p_i) \in \{0,1\}$, eliminating the cognitive burden of granular self-rating scales.

\begin{figure}[htbp]
\centerline{\includegraphics[width=\columnwidth]{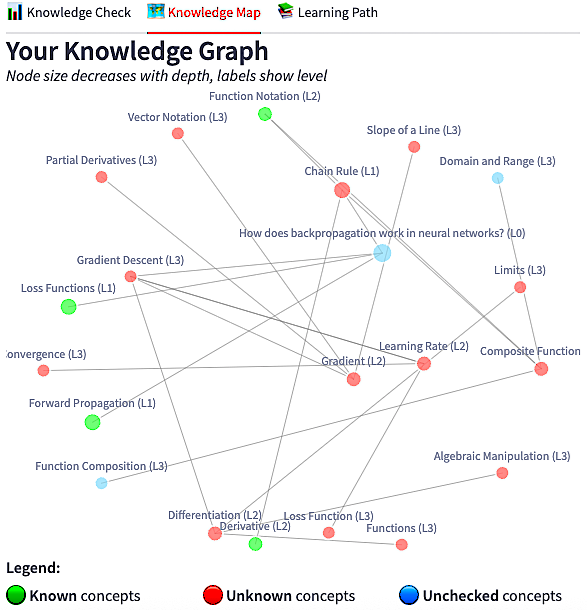}}
\caption{Knowledge dependency graph revealing cross-domain prerequisites across mathematics, programming, and ML/DL domains.}
\label{fig:knowledge_graph}
\end{figure}

When learners mark prerequisites as unknown ($K(p_i) = 0$), the algorithm recursively extracts sub-prerequisites, creating an expanding dependency tree. Recursion continues until reaching either the maximum depth $d_{\max}$ or fundamental concepts requiring no further prerequisites. This bounded approach ensures computational efficiency while discovering multi-level knowledge gaps. The recursive nature addresses the "unknown unknowns" problem by revealing dependencies learners would not anticipate.

\paragraph{Language Model Integration}

GPT-4o integration employs carefully engineered prompts with structured JSON output format for consistent prerequisite extraction. The prompting strategy extracts 2-4 critical prerequisites per concept, a range selected to balance comprehensive coverage with manageable cognitive load. Prompts explicitly instruct the model to identify immediate technical dependencies rather than broad foundational skills, contextualizing extraction based on the specified education level.

For complex concepts, the system implements prompt chaining where initial extraction results inform subsequent queries. This approach enables cross-domain dependency discovery across mathematics, programming, machine learning, and deep learning domains. The structured JSON format ensures consistent parsing while maintaining flexibility across diverse academic topics.

\paragraph{Interactive Interface Design}

The interface design minimizes cognitive load through binary assessment mechanisms replacing traditional multi-point scales. The three-tab structure separates knowledge assessment, visual exploration, and learning path review, allowing focused interaction while maintaining overall progress awareness. Real-time visual feedback immediately reflects assessment decisions—known concepts appear dimmed while unknown concepts expand inline to reveal prerequisites. Level indicators provide hierarchical orientation, and the system handles duplicate concepts intelligently, displaying "Already confirmed" status when prerequisites appear across multiple branches.

\begin{table*}[t]
\centering
\caption{Comparative Analysis of RPKT vs. Standard LLM (GPT-4o) Educational Explanations}
\label{tab:comparison}
\begin{tabular}{p{2.5cm} p{7cm} p{7cm}}
\toprule
\textbf{Criterion} & \textbf{RPKT System} & \textbf{Standard LLM (GPT-4o)} \\
\midrule
\textbf{Knowledge Gap Discovery} & 
Actively discovers "unknown unknowns" through recursive prerequisite tracing without requiring question formulation &
Assumes users can identify and articulate what they need to learn \\
\textbf{User Interaction} & 
Simple binary assessment (know/don't know) eliminates need for question generation &
Requires users to formulate effective questions and prompts to identify gaps \\
\textbf{Cognitive Burden} & 
Minimal: users only make binary decisions, no question formulation needed &
High: users must generate questions about concepts they don't understand \\
\textbf{Pedagogical Strategy} & 
Bottom-up approach; builds from identified knowledge boundaries to target concept &
Top-down approach; explains target concept directly with occasional context \\
\textbf{Content Structure} & 
Dynamic hierarchical tree based on individual assessment results &
Static linear or sectioned format regardless of user knowledge state \\
\textbf{Prerequisite Handling} & 
Systematically identifies and addresses missing prerequisites before main content &
Mentions prerequisites briefly or assumes prior knowledge \\
\textbf{Personalization Level} & 
Personalized based on recursive knowledge state assessment &
Generic explanations with same content for all users \\
\textbf{Learning Path} & 
Generates personalized prerequisite sequence tailored to individual gaps &
Provides standard explanation without customized learning trajectory \\
\textbf{Coverage Completeness} & 
Systematically traces prerequisite dependencies to foundational concepts &
May have gaps in foundational concepts depending on response scope \\
\textbf{Best Use Case} & 
Self-directed learning, knowledge gap identification, comprehensive understanding &
Quick references, specific questions, users with clear learning objectives \\
\textbf{Scalability} & 
Domain-agnostic recursive algorithm designed for cross-subject application &
Depends on training data coverage and prompt engineering quality \\
\bottomrule
\end{tabular}
\end{table*}

\begin{figure}[htbp]
\centerline{\includegraphics[width=\columnwidth]{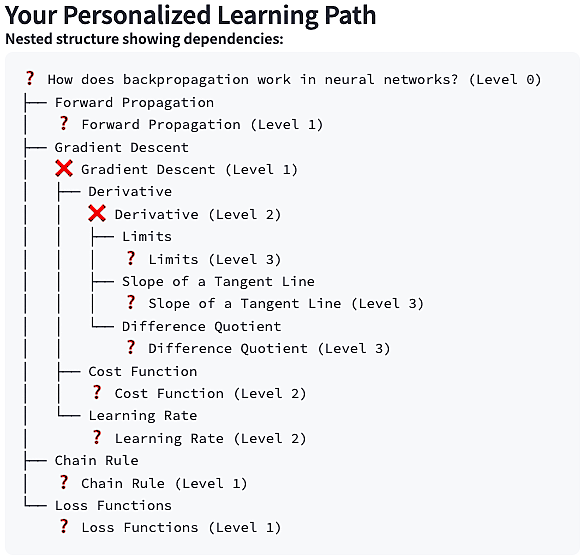}}
\caption{Generated learning path with hierarchical prerequisite structure from foundations to target concept.}
\label{fig:learning_path}
\end{figure}

\section{DEMONSTRATION}

We demonstrate RPKT's capabilities through a case query "How does backpropagation work in neural networks?" to validate our recursive prerequisite discovery approach.

\paragraph{Interface in Practice}
Figure~\ref{fig:main_interface} shows the RPKT system interface presenting the initial analysis for a backpropagation query. The system displays the question understanding, explains why backpropagation is important, and identifies four key Level 1 concepts to check: Forward Propagation, Gradient Descent, Loss Functions, and Chain Rule. This clean interface design with clear sections for question analysis and prerequisite identification demonstrates how the system prepares learners for the assessment phase before any binary decisions are made.

\paragraph{Depth of Discovery}
Figure~\ref{fig:recursive_expansion} demonstrates the recursive expansion mechanism when a user marks "Gradient Descent" as unknown. The system shows "Gradient Descent" displays "Already confirmed" status and expands to reveal its Level 2 prerequisites: Derivative and Cost Function awaiting users' further confirmation. This inline expansion maintains context while revealing the prerequisite hierarchy, validating our approach to discovering knowledge gaps learners cannot identify independently.

\paragraph{Cross-Domain Dependencies}
The knowledge graph in Figure~\ref{fig:knowledge_graph} visualizes the prerequisite network for the backpropagation query. Node sizes decrease with depth (L0 to L3) while colors indicate assessment status: green for known concepts, red for unknown, and blue for unchecked. The graph reveals multi-level dependencies from immediate prerequisites like Gradient Descent (L1) through mathematical foundations like Differentiation (L2) to advanced concepts like Limits (L3), demonstrating the complex prerequisite web that our system systematically uncovers.

\paragraph{Personalized Learning Sequence}
Figure~\ref{fig:learning_path} presents the generated learning path showing nested prerequisites for backpropagation. The hierarchical structure traces dependencies from L0 through L3, with unknown concepts marked with \xmark\ and unassessed concepts marked with ?. For example, Gradient Descent (\xmark) expands to reveal Derivative (\xmark), which further requires Limits (?). This automatically generated sequence guides learners from mathematical foundations through progressive concepts to the target knowledge.

\begin{figure}[!t]
\centering
\includegraphics[width=1\columnwidth]{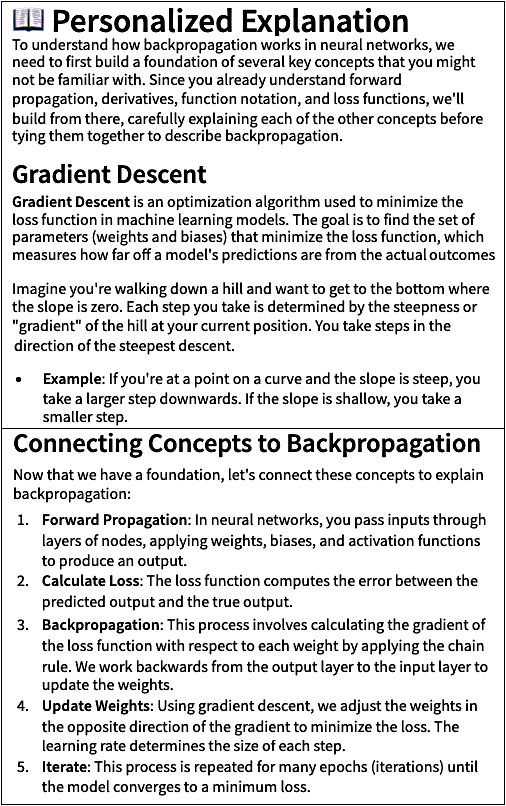}
\caption{Personalized explanation generation based on identified knowledge gaps. Top: System addresses unknown prerequisite (Gradient Descent) while acknowledging existing knowledge. Bottom: Synthesis connecting learned concepts back to the original backpropagation query.}
\label{fig:personalized_explanation}
\end{figure}

\paragraph{Personalized Explanation Generation}
Figure~\ref{fig:personalized_explanation} demonstrate the system's ability to generate tailored explanations based on the identified knowledge gaps. After recursive prerequisite discovery and assessment, RPKT creates a personalized tutorial that explicitly acknowledges the learner's existing knowledge ("Since you already understand forward propagation, derivatives, function notation, and loss functions") and thoroughly explains unknown concepts before connecting them to the target topic. The system first addresses each unknown prerequisite individually, then synthesizes these concepts to explain how they relate to backpropagation. This approach ensures learners build understanding progressively rather than encountering unexplained prerequisites within the main explanation.

\paragraph{Comparative Analysis}
Table~\ref{tab:comparison} contrasts RPKT with standard LLM tutoring approaches. The key distinction lies in knowledge gap discovery: RPKT actively uncovers unknown prerequisites without requiring question formulation, while standard LLMs assume users can articulate their learning needs. RPKT's bottom-up pedagogical strategy builds from identified knowledge boundaries, whereas standard LLMs provide top-down explanations that may assume prerequisite knowledge. This systematic approach addresses the fundamental challenge that learners cannot identify what they don't know they need to learn.

The demonstration confirms several key design choices. The recursive discovery successfully identifies multi-level prerequisite chains that span unexpected domains. The binary assessment interface enables efficient progression without cognitive overload. The visual representations effectively communicate both dependency depth and breadth. These findings validate that dynamic prerequisite discovery through recursive tracing provides a practical solution to the "unknown unknowns" problem in education.

\section{DISCUSSION}

The demonstration of RPKT validates our approach to addressing the "unknown unknowns" problem through recursive prerequisite discovery. The system successfully identifies multi-level prerequisite chains spanning unexpected domains—revealing that understanding backpropagation requires foundational mathematics like calculus and linear algebra that learners would not independently identify.

Our approach is particularly valuable for learners beginning new disciplines or exploring interdisciplinary fields. Traditional learning often overwhelms beginners with complex explanations that assume prerequisite knowledge. RPKT systematically traces back to foundational concepts, preventing the frustration of encountering incomprehensible content. This makes the system especially suited for emerging interdisciplinary domains where prerequisites span multiple fields and traditional curricula may not exist.

The binary assessment interface, visual dependency graphs, and personalized explanations work together to transform learning from "what should I ask?" to simple decisions about current knowledge. By eliminating question formulation and revealing hidden prerequisites, the system enables efficient, comprehensive learning paths.

Currently, we present a demonstration system validating our design approach. The next critical step is experimental testing with real learners to measure learning outcomes and efficiency gains compared to traditional methods. Such empirical evaluation will provide insights into the system's educational effectiveness and guide further refinements.

\section*{Acknowledgment}
Claude refined the writing to improve the overall readability.

\bibliographystyle{IEEEtran}
\bibliography{reference}

\end{document}